\documentclass[prb,twocolumn,showpacs,preprintnumbers,amsmath,amssymb,aps,floats]{revtex4-1}
\usepackage{amssymb}
\usepackage{amsmath}
\usepackage{graphicx,bm}
\usepackage{dcolumn}
\usepackage{psfrag}
\usepackage{color}
\usepackage{verbatim}
\def\be{\begin{equation}}       \def\ee{\end{equation}}
\def\bea{\begin{eqnarray}}      \def\eea{\end{eqnarray}}
\def\bes{\begin{subequations}}  \def\ees{\end{subequations}}
\def\half{\frac{1}{2}}
\def\dag{\dagger}
\def\non{\nonumber}

\def\k{_{\bf k}}
\def\kk{_{-\bf k}}

\begin{document}

\title{Quantum Fluctuation effects on the Ordered Moments in a two dimensional frustrated Ferrimagnet}
\author{Kingshuk Majumdar}
\email{majumdak@gvsu.edu}
\affiliation{Department of Physics, Grand Valley State University, Allendale, 
Michigan 49401, USA}
\author{Subhendra D. Mahanti}
\email{mahanti@msu.edu}
\affiliation{Department of Physics and Astronomy, Michigan State University, East Lansing, 
Michigan 48824, USA}
\date{\today}

\begin{abstract}
\label{abstract}
We propose a novel two-dimensional frustrated quantum spin-1/2 anisotropic Heisenberg model with alternating 
ferromagnetic and antiferromagnetic magnetic chains along one direction and antiferromagnetic interactions along the other. 
The (mean-field) ground state is ferrimagnetic in certain range of the interaction space. Spin-wave theory analysis 
of the reduction of ordered moments at inequivalent spin sites and the instability of the spin waves 
suggest a quantum phase transition which has the characteristics of 
both the frustrated two-dimensional antiferromagnetic S=1/2 ($J_1, J_2$) model and 1D S$_1$=1, S$_2$=1/2 quantum 
ferrimagnetic model.
\end{abstract}
\pacs{71.15.Mb, 75.10.Jm, 75.25.-j, 75.30.Et, 75.40.Mg, 75.50.Ee, 73.43.Nq}

\maketitle

\section{\label{sec:Intro}Introduction}
Low-dimensional quantum spin systems are excellent examples to explore the physics of strongly interacting 
quantum many-body systems.~\cite{diep} In addition to the inherent quantum nature of the interacting elements (for example 
localized spins with S=1/2), these systems provide an array of choices where the effects of competing interactions, 
non-equivalent nearest neighbor bonds, and frustration on quantum fluctuations of the long range order parameter and on
quantum phase transitions at $T=0$K (no thermal fluctuations) can be explored. Although extensive studies using 
different theoretical approaches and using different spin models have been done over the last several 
decades we will first discuss two simple models relevant to our present work.~\cite{mila,subir1}
They are (i) two-dimensional (2D) antiferromagnetic S=1/2 Heisenberg 
model on a square lattice with nearest (NN) and next nearest neighbor (NNN) 
antiferromagnetic interactions ($J_1,J_2$) [Model I] and (ii) one-dimensional (1D) spin chain consisting of alternating 
S$_1$=1 and S$_2$=1/2 spins interacting 
antiferromagnetically [Model II]. The classical ground state (GS) of model I in certain ($J_1, J_2$) domain is long range ordered (LRO) antiferromagnet 
whereas that of model II is a long range ordered ferrimagnet. Quantum spin fluctuations (QSF) dramatically affect the physical properties of 
these systems, which we review briefly after first introducing a new model [Model III] below. 

We propose a novel 2D Heisenberg model at $T=0$K consisting of 
only S=1/2 spins, which combines the essential features of the two above models, extreme anisotropy of the NN 
bonds (some ferro and some antiferro) and frustration. The classical ground state (discussed in detail later in the
paper) is a four-sublattice ferrimagnet 
in certain parameter space. Our focus in this paper is on the stability of this ferrimagnetic ground state and effect of 
QSFs at $T=0$K on the 
long range ordered sublattice magnetizations.

\section{\label{sec:Models}Review of Models I and II}
\subsection{Model I} The classical ground state of the 2D S=1/2 ($J_1, J_2$) Heisenberg model on a square lattice depends on the 
frustration parameter $\eta=J_2/J_1$.\cite{diep,mila} 
For $\eta <0.5$, the GS is a Ne\'{e}l state with ordering wave vector ($\pi,\pi$), similar to the unfrustrated case 
whereas for $\eta>0.5$ the GS is 
the degenerate columnar antiferromagnetic state (CAF) with ordering wave vectors ($\pi,0$) and ($0,\pi$). There is 
a first-order phase 
transition from the Ne\'{e}l state to CAF state at $\eta=0.5$. Effects of QSF on this phase transition and other properties of this 
model have been investigated using a large number of 
methods.~\cite{anderson,harris71,igar93,majumdar10,richter,bishop,sandvik01,isaev,syro,kivelson} Here we 
review the main results obtained within linear spin wave theory (LSWT). Sublattice magnetization, $m$ is reduced 
by QSF  from its classical value of 0.5 to 0.303 at $\eta=0$ and then decreases 
monotonically with increasing $\eta$ and approaches zero at the first critical point $\eta_{c1}=0.38$. Similarly
$m=0.37$ at $\eta=1$ and then steadily decreases to zero at the second critical point $\eta_{c2}=0.49$. 
LSWT clearly indicates that QSF effects are enhanced in the presence of frustration. Also it suggests that in 
the region $\eta_{c1} <\eta<\eta_{c2}$, the classical GSs are not stable. The nature of the GS (e.g. spin-liquid, valence bond) 
and low energy excitations in this region
have been extensively studied during past several years and continue to be of great current interest.

\subsection{Model II} The second model deals with ferrimagnets. Ferrimagnets are somewhere between ferromagnets and 
antiferromagnets.~\cite{mikeska,ivanova,Kolezhuk1,brehmer,pati1,pati2,NBIvanov0,Kolezhuk2,shoji1,shoji2,NBIvanov1,NBIvanov3}
It is well known that for 1D quantum S=1/2 ferromagnet, the ground state is long-range ordered 
and QSFs do not reduce the classical value of $m$. In contrast, in a 1D quantum S=1/2 antiferromagnet (AF), QSFs 
completely destroy the classical LRO. The question what happens for ferrimagnets drew considerable interest 
in the late 90's and several interesting works were done using a simple isotropic NN antiferromagnetic 
Heisenberg model with two types of spins, ${\rm S}_1=1$ and ${\rm S}_2 = 1/2$ in a magnetic 
unit cell (MUC).~\cite{Kolezhuk2, brehmer, pati1, NBIvanov0, NBIvanov1, NBIvanov2, NBIvanov3,shoji1,shoji2}
Following Refs.~[\onlinecite{brehmer}] and [\onlinecite{Kolezhuk2}] we discuss some 
of the interesting physical properties of this 1D system and point out how our proposed 2D model differs from this.

The Hamiltonian of the 1D system is given by
\be
{\cal H}=\sum_n\left[ J\big({\bf S}_{1n}\cdot {\bf S}_{2n}+{\bf S}_{2n}\cdot {\bf S}_{1n+1}\big)
-hS_{Tn}^z\right],
\ee
where ${\bf S}_{1n}$ and ${\bf S}_{2n}$ are spin-1 and spin-1/2 operators respectively in the $n^{\rm th}$ unit cell (UC), effective 
field $h(=g\mu_B H$ with
$g$ gyro-magnetic ratio, $\mu_B$ Bohr magneton, and $H$ external magnetic field) and $S_{Tn}^z=S_{1n}^z+S_{2n}^z$.

According to the Lieb-Mattis theorem~\cite{Lieb}, for $H = 0$, the GS is long range ordered as 
the system has total spin $S_T=N/2$, where $N$ is the number of UCs in GS, $\langle S_{1n}^z \rangle=1$ and 
$\langle S_{2n}^z \rangle=0.5$. The problem of looking at the excitations is well suited
for the LSWT approach. Since the elementary magnetic cell consists of two spins, LSWT predicts 
two types of magnons: a gapless ``acoustic'' or ``ferromagnetic'' branch with $S_{T}^z=N/2-1$, and a 
gapped ``optical'' or ``antiferromagnetic'' branch with $S_{T}^z=N/2+1$. The optical magnon gap for this model 
has been numerically found to be $\Delta_{\rm opt}=1.759J$.~\cite{shoji2}
An intriguing property of this 1D  quantum ferrimagnet is that when one turns on the magnetic field $H$, 
the acoustic branch opens up a gap but the optical gap decreases and at a critical value of the field $H_{c1}$ this gap 
vanishes, the system then enters into a quantum spin liquid (QSL) phase, where the GS is dominated by QSFs with 
spinon-like excitations.~\cite{Kolezhuk2,brehmer,pati1} With further increase in the strength of the field this QSL phase 
goes into a saturated ferromagnetic phase.

Brehmer et al.~[\onlinecite{brehmer}] calculated the sublattice magnetization for the S=1 sublattice ($m_A$) and found 
it to be (1-$\tau$) with $\tau \approx 0.305$. The sublattice magnetization of 
the S=1/2 sublattice can be calculated using their method and is found to be $m_B=-0.5+\tau$. The ordered moment 
of the S=1/2 sublattice is reduced by a factor of $\sim 2.5$ due to QSF. There are 
two important points worth noting here: (1) the total magnetization (ferromagnetic) per magnetic unit cell is
$m_A+m_B=0.5$, the classical value and (2) QSF reduction of the S=1/2 sublattice is larger than the 2D S=1/2 Heisenberg model 
for a square lattice where $\eta \sim 0.2$. Point (1) is consistent with the fact that the ferromagnetic long range order is 
not affected by QSF. Also $m_A$ and $m_B$ are independent of the magnetic field.

\section{Model III} 
\begin{figure}[httb]
\centering
\includegraphics[width=2.0in,clip]{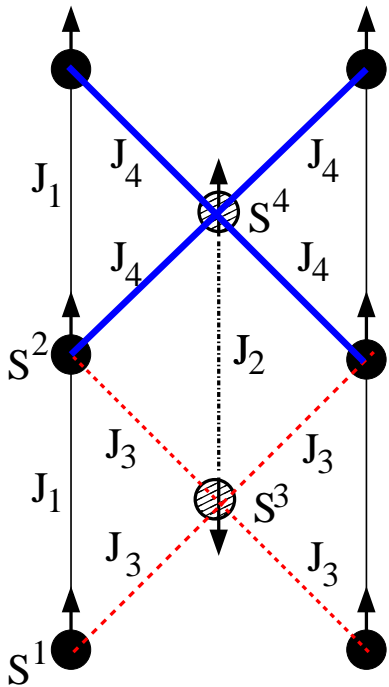}
\caption{\label{fig:CrMWstruc1}(Color online) 
Classical ferrimagnetic ground state 
of 2D F-AF chain. The basic magnetic unit cell comprises of three up-spins
${\bf S}_1, {\bf S}_2, {\bf S}_4$ and one down-spin ${\bf S}_3$. The interaction strengths
$J_1, J_2, J_3$ are all positive and $J_4$ is the frustrated bond.}
\end{figure}

As mentioned in the beginning, we introduce a new 2D Heisenberg model which incorporates different aspects of the two 
models discussed above, anisotropic bonds and frustration. Also, instead of two types of spins and single exchange 
parameter, our model consists of only S=1/2 spins interacting with Heisenberg exchange couplings of different 
signs (both ferro and antiferro). The unit cell consists of four types of spins which we denote as ${\bf S}^{(\mu)} \;(\mu=1..4)$,
it is a Bravais lattice. The lattice vectors for the four spins in 
a rectangular lattice with parameters ($a,b$) along the 
$x$ and $y$ directions are given by ${\bf R}_{i\mu}=i_x a{\bf \hat x}+i_y b{\bf \hat y}+{\bm \tau}_{\mu}$ where
${\bm \tau}_1=(0,0), 
{\bm \tau}_2=(0,b/2), {\bm \tau}_3=(a/2,b/4)$ and ${\bm \tau}_4=(a/2,3b/4)$ (see Fig.~\ref{fig:CrMWstruc1}). 
As we will show, the ground state is ferrimagnetic in certain range of 
exchange  parameter space. Three spins combine to form the S=3/2 sublattice. In contrast to the 1D 
S=(3/2,1/2) model, where the magntitudes of the spins in each sublattice are fixed, in our model, the 
S=3/2 sublattice can undergo amplitude fluctuations. In fact, the present model was inspired by recent inelastic neutron scattering experiments 
on a quasi 2D spin systems containing Cu$^+_2$ ions, Cu$_2$(OH)$_3$Br.~\cite{Ke} However, in this system the effect 
of orbital ordering of active magnetic orbitals driven by the ordering of the Br$^+$ ions on the exchange parameters 
is such that the ground state is an antiferromagnet with eight spins per unit cell.

The Heisenberg spin Hamiltonian (${\cal H}$) for model III is divided into two parts, intra-chain (${\cal H}_1$) and 
inter-chain (${\cal H}_2$):  
\be 
{\cal H}={\cal H}_1 + {\cal H}_2,
\label{fullH}
\ee
where
\begin{subequations}
\label{ham}
\bea
{\cal H}_1 &=& -J_1\sum_i \Big[{\bf S}_i^{(1)}\cdot {\bf S}_i^{(2)}+
\frac 1{2}\Big({\bf S}_i^{(1)}\cdot {\bf S}_{i-b\hat y}^{(2)}+ {\bf S}_i^{(2)}\cdot {\bf S}_{i+b\hat y}^{(1)}\Big) \Big] \non \\
&+& J_2\sum_i \Big[{\bf S}_i^{(3)}\cdot {\bf S}_i^{(4)}+
\frac 1{2}\Big({\bf S}_i^{(3)}\cdot {\bf S}_{i-b\hat y}^{(4)}+ {\bf S}_i^{(4)}\cdot {\bf S}_{i+b\hat y}^{(3)}\Big) \Big],\non \\
\\
{\cal H}_2 &=& \half J_3\sum_i \Big({\bf S}_i^{(1)}+{\bf S}_i^{(2)}\Big)
\cdot \Big({\bf S}_i^{(3)}+{\bf S}_{i-a\hat x}^{(3)}\Big)\non \\
&+& \half J_4 \sum_i \Big[{\bf S}_i^{(1)}\cdot \Big({\bf S}_{i-b\hat y}^{(4)}+{\bf S}_{i-a\hat x-b\hat y}^{(4)}\Big)\non \\
&+&{\bf S}_i^{(2)}\cdot \Big({\bf S}_i^{(4)}+{\bf S}_{i-a\hat x}^{(4)}\Big)\Big]\non \\ 
&+& \half J_3 \sum_i {\bf S}_i^{(3)}\cdot \Big({\bf S}_i^{(1)}+{\bf S}_{i+a\hat x}^{(1)}+{\bf S}_i^{(2)}
+{\bf S}_{i+a\hat x}^{(2)}\Big) \non \\
&+& \half J_4 \sum_i{\bf S}_i^{(4)}\cdot \Big({\bf S}_i^{(2)}+{\bf S}_{i+a\hat x}^{(2)}+{\bf S}_{i+b\hat y}^{(1)}
+{\bf S}_{i+a\hat x+b\hat y}^{(1)}\Big).\non \\
\eea
\end{subequations}
All exchange parameters $J_\mu $ are positive (see Fig.~\ref{fig:CrMWstruc1} for an illustrative 
long range ordered ferrimagnetic). We refer to this model as ($J_1, J_2, J_3, J_4$) model. 

{\em Classical Ground State:} The basic model consists of alternating 1D ferro (strength $J_1=1$) and 
antiferromagnetic (strength $J_2=\eta_2 J_1$) S=1/2 chains (along the $y$-axis). The nearest chains interact with 
interaction strengths $J_3$ ($= \eta_3 J_1$) and $J_4$ ($= \eta_4 J_1$) which are 
antiferromagnetic. Before discussing the excitations and quantum spin fluctuations, we first consider the
ground state 
of our model when the spins are treated classically (mean field state). With $J_3=J_4=0$, the ground state ($G_0$)
with broken global symmetry consists of decoupled alternating ordered F chains ({\bf S}$_1$ and {\bf S}$_2$ spins) and 
AF chains ({\bf S}$_3$ and {\bf S}$_4$ spins). Due to the time reversal symmetry, the F chains can be either up or
down (chosen arbitrarily) and the AF chains can be in one of the two Ne\'{e}l states. The degeneracy of the $G_0$ is $2^{2M}$,
where $M$ is the number of F (or AF) chains. For $J_3>0$ and $J_4=0$, if we fix the orientation of one F chain, the nearest 
two AF chain orientations are fixed by the $J_3$ bond. The neighboring F chain orientations are then fixed. In this way, we 
have the exact ground state $G$ as each bond takes its minimum energy value. When $\eta_3>0$ and $\eta_4<0$ (ferromagnetic),
the system is not frustrated and the classical GS is 
a collinear ferrimagnetic state as shown in Fig.~\ref{fig:CrMWstruc1}. 
However, for $\eta_3>0$ and $\eta_4>0$, spin ${\bf S}_4$ is frustrated. For
weak frustration i.e. $\eta_4<<\eta_3$, $G$ is most likely the exact ground state and with increasing frustration ($J_4$)
the system will undergo a phase transition to a new state which may or may not be long range ordered. One approach to attack the problem 
is to use the generalized Luttinger Tisza method [\onlinecite{LT}] first proposed by Lyons and Kaplan.\cite{LK} It turns out that
for our Bravais lattice with four-spin/unit cell system the calculations are quite difficult. So in the absence of the 
knowledge of the exact ground state for large $J_4$, we have used a different approach. We study the local stability of $G$
with increasing strength of the frustrating bond ($J_4$). As we will show later, depeding on the strength of $J_2/J_1$, there
is a critical value of $J_4/J_3$ where the ground state $G$ is no longer locally stable. Thus in our current analysis of the
phase diagram and excitations of the model using spin-wave theory we use $G$ as the ground state.



\section{Spin-wave Theory}
It is well-known that spin-wave theory is best suited to treat the dynamics of long range-ordered states in quantum
spin system with large spin $S$. In the leading order (linear spin wave theory - LSWT), the excitations are magnons. 
When magnon-magnon interaction effects are negligible (for example for $S>>1/2$ and three dimensions), LSWT provides a 
very good description of the quantum spin fluctuation effects, one example being the reduction of the ordered moment in 
Heisenberg quantum antiferromagnets. However, for S=1/2 systems in 2D, magnon-magnon interactions are not negligible 
and one must incorporate higher order spin (1/S) corrections to treat the system.\cite{igar92,igar93,igar05,majumdar10} Even for these systems, 
LSWT provides 
qualitatively correct physics. For example, for 2D Heisenberg spin systems with nearest neighbor (NN) antiferromagnetic (AF) 
coupling [($J_1,J_2$) model with no frustration i.e. $J_2=0$] on a square lattice, the ordered moment (average sublattice spin $\langle S_z \rangle$) reduces due to 
QSF from 0.5 to 0.303 as given by LSWT.\cite{anderson,harris71} When one includes the higher order magnon-magnon interaction 
effects using (1/S) expansion theory  $\langle S_z\rangle = 0.307$,\cite{igar05,majumdar10} indicating that LSWT is very reasonable. 
For the general ($J_1,J_2$) model, the effect of frustration is much more subtle. Frustration tends to destabilize long range order. 
With
increase in the strength of frustration, $\langle S_z \rangle = 0$ at a critical value of $J_2= J_{2c}$. LSWT 
gives $J_{2c}=0.38$ whereas including the magnon-magnon interaction one finds $J_{2c}=0.41$,\cite{igar93,majumdar10} again indicating the 
reasonableness of LSWT in providing a measure of the QSF induced reduction of the magnetization $M$. 
In a recent work (Ref.~[\onlinecite{syro}]) results for this model is obtained using a four-spin 
bond operator technique where it is found that $\langle S_z \rangle = 0.301$ for $J_2=0$ and $J_{2c} = 0.36$, which are close 
to the LSWT results. We should mention here that all these method fail in the spin disordered state i.e. when 
$J_2 > J_{2c}$. 

In view of the above discussion, we opted to use LSWT to analyze the effect of QSF on the average magnetic moment and the 
critical strength of the frustration where the ordered moments vanish. Unlike the ($J_1,J_2$) model (two sublattice with same
value of the ordered moment) our 2D frustrated ($J_1, J_2, J_3, J_4$) model has a 4-sublattice structure as shown below 
and different sublattice moments are affected differently by QSF.

For our analysis we only
consider the parameter space $(\eta_2,\eta_3,\eta_4)$ of the Hamiltonian ${\cal H}$ [Eq.~\eqref{fullH}] where the 
GS is stable and is long range ordered collinear ferrimagnetic state. 
The spin Hamiltonian in Eq.~\eqref{ham} is mapped onto a Hamiltonian of interacting bosons by expressing the spin operators in 
terms of bosonic creation 
and annihilation operators $a^\dag, a$ for three ``up'' spins (spins 1, 2, and 4) and $b^\dag, b$ for one ``down' spin (spin 3) 
using the standard Holstein-Primakoff representation~\cite{HP}
\begin{eqnarray}
S_{in}^{+ i} &\approx& \sqrt{2S}a_{in},\;
S_{in}^{- i} \approx \sqrt{2S}a_{in}^{\dag},\;
S_{in}^{z i} = S-a^{\dag}_{in} a_{in}, \label{hol1} \non \\ 
S_{jn}^{+ j} &\approx& \sqrt{2S}b_{jn}^\dag,\;
S_{jn}^{- j} \approx \sqrt{2S}b_{jn}, \;
S_{jn}^{z j} = -S+b^\dag_{jn}b_{jn}, \label {hol2}\non 
\label{holstein}
\end{eqnarray}
and expand the Hamiltonian [Eq.~\eqref{ham}] perturbatively in powers of $1/S$ keeping terms only up to 
the quadratic terms. The resulting quadratic Hamiltonian is given as:
\be
{\cal H}=E_{\rm cl}+{\cal H}_0 + \cdots, 
\ee
where 
\be
E_{\rm cl} = -2J_1NS^2\big[1+\eta_2+2(\eta_3-\eta_4)\big]
\label{classH}
\ee
is the classical GS energy and 
\bea
&&{\cal H}_{0}= 2SJ_1\sum_{{\bf k}\in {\rm BZ}}\Big[(1+\eta_3-\eta_4)\Big(a^{(1)\dag}\k a\k^{(1)}+ 
a^{(2)\dag}\k a\k^{(2)}\Big)\non \\
&&-\gamma_y\Big(a^{(1)}\k a\k^{(2)\dag}+ a^{(1)\dag}\k a\k^{(2)} \Big)
+ (\eta_2-2\eta_4)a^{(4)\dag}\k a\k^{(4)} \non \\
&&+ (\eta_2+2\eta_3)b^{(3)\dag}\kk b\kk^{(3)} 
+\eta_2\gamma_y\Big(b^{(3)\dag}\kk a\k^{(4)\dag}+ b^{(3)}\kk a\k^{(4)} \Big) \non \\
&&+\eta_3\gamma_x\Big(e^{ik_yb/4}a^{(1)\dag}\k b\kk^{(3)\dag}+ e^{-ik_yb/4}a^{(2)\dag}\k b\kk^{(3)\dag} +h.c.\Big)\non \\
&&+\eta_4 \gamma_x\Big(e^{-ik_yb/4}a^{(1)\dag}\k a\k^{(4)}+ e^{ik_yb/4}a^{(2)\dag}\k a\k^{(4)} +h.c.\Big)
\Big]\label{H0}
\eea
with $\gamma_x=\cos (k_xa/2)$ and $\gamma_y=\cos (k_y b/2)$. 

In the absence of inter-chain coupling ($\eta_3=\eta_4=0$) the magnon spectrum can be obtained using the standard 
Bogoliubov transformations.~\cite{Bogo} We find
four modes for each $k_y\;(-\pi/b <k_y<\pi/b)$ independent of $k_x\;(-\pi/a<k_x<\pi/a)$: two from the F-chains ($\alpha$-branches) 
and two from the AF-chains (one $\alpha$ and one $\beta$). The quadratic Hamiltonian takes the following form:
\begin{eqnarray}
{\cal H}_0 &=& \sum_{{\bf k}\in {\rm BZ}} \Big[\epsilon\k^{(1)} \alpha^{(1)\dag}\k\alpha\k^{(1)}+ 
\epsilon\k^{(2)} \alpha^{(2)\dag}\k\alpha\k^{(2)} \non \\
&+& \epsilon\k^{(3)} \Big(\alpha^{(4)\dag}\k\alpha\k^{(4)}+ 
\beta^{(3)\dag}\kk \beta\kk^{(3)}\Big) \Big]
+ \sum_{{\bf k}\in {\rm BZ}} \Big(\epsilon\k^{(3)}-2J_1S \Big),\non \\
\label{H0special}
\end{eqnarray}
where
\begin{subequations}
\label{spcase}
\begin{eqnarray}
&&\epsilon\k^{(1,2)}= 2J_1S[1 \mp \gamma_y], \\
&&\epsilon\k^{(3)}= 2J_2S\sqrt{1 - \gamma_y^2}=2J_2S|\sin (k_y b/2)|.
\end{eqnarray}
\end{subequations}
The last term in Eq.~\eqref{H0special} are the LSWT corrections to the classical ground state energy $E_{\rm cl}$ 
in Eq.~\eqref{classH} for the special case $\eta_3=\eta_4=0$.

With inter-chain coupling (i.e. $\eta_3, \eta_4>0$), we have not been able to find the analytical Bogoliubov transformations that 
transforms the bosonic spin operators to Bogoliubov quasiparticle operators that 
diagonalize the Hamiltonian ${\cal H}_0$ [Eq.~\eqref{H0}]. For the special case $k_x=\pi/a$ i.e. $\gamma_x=0$, 
we use the equation of motion method (see Appendix \ref{EMM}) and obtain analytical solutions for the 
magnon dispersion which are:
\begin{subequations}
\bea
&&\epsilon\k^{(1,2)}=2J_1S\big[(1+\eta_3-\eta_4) \pm \gamma_y\big], \label{dispgx0F} \\
&&\epsilon\k^{(3,4)}=2J_1 S\big\vert(\eta_3+\eta_4) \pm \sqrt{(\eta_3-\eta_4+\eta_2)^2-\eta_2^2\gamma_y^2}\big \vert.\non \\
\label{dispgx0AF}
\eea
\label{anaeqs}
\end{subequations}
When $\eta_3=\eta_4=0$ the above dispersions reduce to Eq.~\eqref{spcase} as expected.

For the general case we use an elegant method developed by Colpa to obtain both the eigenenergies 
(magnon dispersions) and eigenvectors (required for the calculation of magnetization).\cite{Colpa,TL} First we write the $8 \times 8$ Hamiltonian [Eq.~\eqref{H0}] in a symmetrized form:
\begin{eqnarray}
{\cal H}_{0} &=& J_1S\sum_{{\bf k}\in {\rm BZ}} \sum_{i=1}^8 X\k^{(i)\dag}h\k X\k^{(i)} \non \\
&-& 2J_1SN\left[1+\eta_2+2(\eta_3-\eta_4)\right],
\end{eqnarray}
with the eigenvectors \\
$X\k=[a\k^{(1)}, a\k^{(2)}, a\k^{(4)}, b\k^{(3)}, a\k^{(1)\dag}, a\k^{(2)\dag}, a\k^{(4)\dag}, b\k^{(3)\dag}]$. 
The hermitian matrix $h\k$ is:
\be
h\k=
\begin{bmatrix} A_1 & -B_1 & C_2^* & 0 & 0 & 0 &0 &C_1 \\ 
-B_1 & A_1  & C_2 & 0 & 0 & 0 &0 & C_1^* \\
C_2 & C_2^* & A_2  & 0 & 0 & 0 &0 & B_2 \\
0 & 0 & 0 &A_3 & C_1 & C_1^* & B_2 & 0\\
0 & 0 & 0 &C_1^* & A_1 & -B_1 & C_2 & 0 \\
0 & 0 & 0 &C_1 & -B_1 & A_1 & C_2^* & 0 \\
0 & 0 & 0 &B_2 & C_2^* & C_2 & A_2 & 0 \\
C_1^* & C_1 & B_2 & 0 & 0 & 0 & 0 & A_3
\end{bmatrix},
\label{hkmatrix}
\ee
where the constants are given in Eqs.~\eqref{coeffs}. 

The Cholesky decomposition has to be applied on $h\k$ to find the complex $K$ matrix that fulfills the condition $h\k=K^{\dag}K$.
However, the Cholesky decomposition only works if the matrix $h\k$ is positive definite (i.e. the eigenvalues are all 
positive).~\cite{Colpa} In case the spectrum of the Hamiltonian ${\cal H}_0$ contains zero modes, one can add a small positive 
value to the diagonal of $h\k$ to make the matrix positive ``definite''. 
We find that the criterion for the Cholesky decomposition to work for all ${\bf k}$ is
$\eta_4 \le \eta_2\eta_3/(\eta_2+2\eta_3)$. As an example, with $\eta_2=3.0, \eta_3=0.4$, 
$\eta_4 \le \eta_{4c}$, where $\eta_{4c}=0.316$. If $\eta_4>\eta_{4c}$ the matrix $h\k$ is not positive
definite and the procedure fails. As we discuss later, this is precisely the same condition for the stability of 
the ferrimagnetic state.
After obtaining the matrix $K$, we solve the eigenvalue problem of the 
hermitian matrix $KgK^{\dag}$, where $g$ is a diagonal paraunitary 
matrix with elements $g_{ii}={\rm diag}(1, 1, 1, 1, -1, -1, -1, -1)$. The resulting eigenvectors are then
arranged in such a way that the first four diagonal elements of the diagonalized $L=U^\dag KgK^\dag U$ matrix are positive and 
the last four elements are negative. The first four positive diagonal elements correspond to the magnon dispersions.

To calculate the sublattice magnetization $m_i$ we first construct the diagonal matrix, $E=gL$ and then find 
the transformation matrix $T$, which relates the boson modes $X\k$ with the Bogoliubov modes $\alpha\k$ via $X\k=T\alpha\k$. 
The matrix $T$ is calculated using~\cite{TL}:
$
T=K^{-1}UE^{1/2}.
$
$m_{i=1,2,4}$ of spins ${\bf S_1}, {\bf S_2}, {\bf S_4}$ are positive but $m_3$ for spin ${\bf S_3}$ is negative.
So we calculate the magnitude of $m_{i=1-4}$ for each of the four sublattices using
\be
\vert m_i \vert =0.5-\vert \tau_i \vert.
\ee
where $\tau_i$ are the reduction caused by QSFs:
\be
\vert \tau_i \vert=\frac 1{N} \sum_{{\bf k}\in {\rm BZ}}\Big\{T\k {\cal D} T\k^\dag\Big\}_{i+4,i+4}.
\ee
${\cal D}$ is a diagonal matrix with $[0, 0, 0, 0, 1,  1, 1, 1]$ as the diagonal elements. We again reiterate that the parameters
$\eta_2, \eta_3, \eta_4$ are chosen such a way that the condition for the Cholesky decomposition is 
satisfied, i.e. $\eta_4 \le \eta_{4c}$.  

\section{\label{sec:results}Magnon Dispersion and Sublattice Magnetization}
\subsection{Magnon Dispersion} 

\begin{figure}[httb]
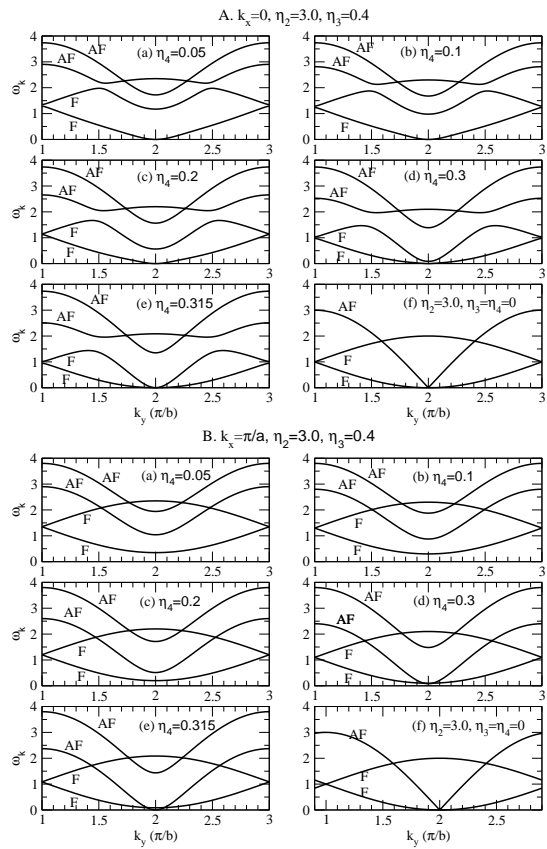

\centering
\includegraphics[width=2.8in,clip]{Dispersionkx0J23J304eta4.eps}
\qquad 
\includegraphics[width=2.8in,clip]{DispersionkxPiJ23J304eta4.eps}
\caption{Magnon dispersion of the ferrimagnetic state for A. $k_x=0$ and B. $k_x=\pi/a$ [Figs. (a-e)] 
with $\eta_2=3.0, \eta_3=0.4$. The frustration 
parameter $\eta_4$ is varied from 0.05 (small frustration) to $\eta_{4}=0.315$.
(f) Limiting case with no inter-chain coupling: the two AF-branches are degenerate, the F-branches 
are gapped, and the lower F-branch vanishes at $k_y=2\pi/b (=0)$. Note that due to 2$\pi$ periodicity 
$k_y=[-\pi/b, \pi/b]=[\pi/b, 3\pi/b]$.}
\label{fig:energydispcomp}
\end{figure}

Effects of inter-chain interaction on the magnon dispersion is displayed in Fig.~\ref{fig:energydispcomp}(a-e) 
where for illustration we have chosen $\eta_2=3, \eta_3=0.4$
and the frustration parameter $\eta_4$ is increased from 0.05 to 0.315. The dispersion along $k_y$ (along the chains) 
is given for two values of $k_x$: 
$k_x= 0$ (top two panels) and $k_x=\pi/a$ (bottom two panels). Also for comparison we give the dispersions for the 
non-interacting chains ($\eta_3=\eta_4=0$). Later we will discuss the $k_x$ dependence 
for some special modes. As expected, 
there are four magnon modes for each ${\bf k}$. For the non-interacting chains, there are two F-magnon modes which 
are split (the lower mode $\sim k_y^2$ for small $k_y$) and two AF-magnons which are degenerate ($\sim k_y$ for small $k_y$). 
In the presence of couplings (discussed below) we will (loosely) refer to these four modes as two F and two AF modes. 

\begin{figure}[httb]
\centering
\includegraphics[width=2.0in,clip]{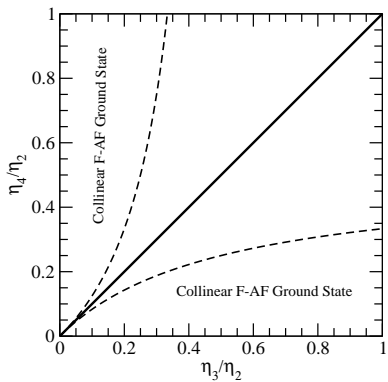}
\caption{Phase diagram of ${\cal H}$ [Eq~\eqref{fullH}]: normalized ${\tilde \eta_4}=\eta_4/\eta_2$ is plotted 
against normalized ${\tilde \eta_3}=\eta_3/\eta_2$. The dashed lines are given by the equations 
${\tilde \eta_4}={\tilde \eta_3}/(1+2{\tilde \eta_3})$ (lower one) 
and ${\tilde \eta_3}={\tilde \eta_4}/(1+2{\tilde \eta_4})$ (upper one). 
They are the boundaries of the stability of the ferrimagnetic
state. The solid thick line ${\tilde \eta_4}={\tilde \eta_3}$ is most likely a critical line.}
\label{fig:phaseD}
\end{figure}

First we consider the case $k_x=\pi/a$ (bottom two panels) where the hybridization between the F and AF modes is absent (as $\gamma_x=0$) - so  
the F and AF chains interact only through effective fields. In this limit, we find from Eq.~\eqref{dispgx0F} and 
Eq.~\eqref{dispgx0AF} that the F-modes get rigidly shifted upwards by  
$2J_1S(\eta_3-\eta_4)$, the two degenerate AF-modes 
are split by $4J_S(\eta_3+\eta_4)$, and both the modes $\sim k_y^2$. At $k_y=0$ the lower F-mode and the lower AF-mode
are gapped, $\Delta_{\rm F}(\pi/a,0)=2J_1S(\eta_3-\eta_4)$ and  
$\Delta_{\rm AF}(\pi/a,0)=2J_1S[\sqrt{(\eta_2+\eta_3-\eta_4)^2-\eta_2^2}-(\eta_3+\eta_4)]$. When the 
frustration parameter $\eta_4$ is increased towards $\eta_3$, there is a
critical value $\eta_{4c}=\eta_2\eta_3/(\eta_2+2\eta_3)<\eta_3$, where  $\Delta_{\rm AF}(\pi/a,0)=0$ 
but $\Delta_{\rm F}(\pi/a,0) >0$. The ferrimagnetic GS  becomes locally unstable and the system 
transits to a new ground state (For the parameter values we have chosen $\eta_{4c}=0.316$ - this is also the place where 
Cholesky decomposition fails because the matrix $h_{\bf k}$ is not positive definite). This is similar to the field 
induced quantum phase transition as a function of the external magnetic field for the 1D quantum  
${\rm S}_1=1, {\rm S}_2=1/2$ model discussed in the introduction.\cite{brehmer,Kolezhuk2} Here the optic mode gap goes to zero at a critical 
field and the system undergoes a quantum phase transition from a ferrimagnetic state to some other state. This 
phase transition occurs in the range $\eta_3>\eta_4>\eta_{4c}$. Fig.~\ref{fig:phaseD} shows a schematic phase diagram in the
$(\eta_4/\eta_2,\eta_3/\eta_2)$ space. 
We also note that for given  $\eta_3$ and $\eta_4 \le \eta_3$, the strength of the exchange in the AF chains $\eta_2$ 
should be greater than a 
critical value $\eta_{2c}=2\eta_3\eta_4/(\eta_3-\eta_4)$ for the ferrimagnetic state to be stable.

\begin{figure}[httb]
\centering
\includegraphics[width=3.2in,clip]{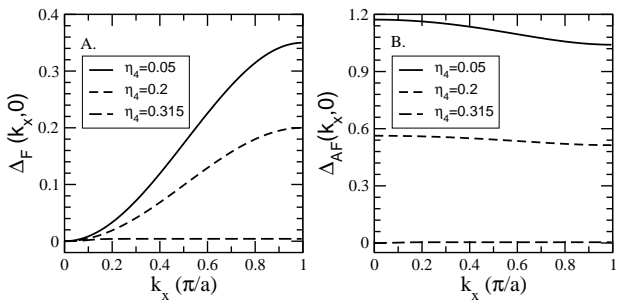}
\caption{Gaps for F-mode ($\Delta_{\rm F}$) and AF-mode ($\Delta_{\rm AF}$) with increase in $k_x$ for $k_y=0$ with
$\eta_2=3.0, \eta_3=0.3$ and three different values of $\eta_4=0.05, 0.2,$ and 0.315. 
}
\label{fig:gap}
\end{figure}

For $k_x=0$, the picture is qualitatively similar, but with two fundamental differences resulting from hybridization between 
ferro and antiferro chain excitations. First, the lower F-mode goes 
to zero when $k_y \rightarrow 0$ as it should for the Goldstone mode. However the dispersion for large  $k_y$ differs 
qualitatively from the 
non-interacting chains. Second, hybridization between the upper F-mode and the lower AF-mode opens up 
a hybridization gap at a finite $k_y$ and the size of the gap increases with $\eta_4$. However, as for the $(k_x,k_y)=(\pi/a,0)$
the gap $\Delta_{\rm AF}(\pi/a,0) \rightarrow 0$ as $\eta_4 \rightarrow \eta_{4c}$. 
In fact $\Delta_{\rm AF}(k_x,0) \rightarrow 0$  for all values of $0\le k_x\le \pi/a$ for $\eta_4 \rightarrow \eta_{4c}$.
In Fig.~\ref{fig:gap}B we show the  $k_x$ dependence of $\Delta_{\rm AF}(k_x,0)$ for three different values of the 
frustration parameter $\eta_4$. Also we show in Fig.~\ref{fig:gap}A the  $k_x$ dependence of $\Delta_{\rm F}(k_x,0)$. 
This suggests that the chains become dynamically decoupled and since the decoupled AF chains are spin liquids without any long range
order, the system goes from an ordered state to a spin disordered state when $\eta_4>\eta_{4c}$. Exact calculations will tell
us about the precise nature of the ground state for $\eta_{4c} <\eta_4<\eta_3$.

\subsection{Sublattice Magnetization}
Following Colpa's method we have calculated the sublattice magnetizations $m_i$ for the four sites.
We have checked that the sum of the reduction in the four sublattice moments due to quantum 
fluctuations, $\sum_{i=1}^4 \tau_i = 0$, which results in the total
magnetic moment equal to one as expected. This is equivalent to the results obtained for S$_1=1$, S$_2=1/2$ 1D quantum
ferrimagnetic state for which the total magnetization/unit cell is equal to 0.5.
Next we discuss the effect of frustration on the quantum fluctuation induced reduction of the 
long-range ordered moments for the four different spins of the unit cell. In 
the absence of interchain coupling [Fig.~\ref{fig:CrMWstruc1}], 
$m_1=\langle S_{1z} \rangle =m_2=\langle S_{2z} \rangle =0.5$ and 
$m_3=\langle S_{3z} \rangle = -m_4=\langle S_{4z} \rangle =0$ (due to quantum spin fluctuation in 1D AF). 
When we turn on $\eta_3$, its effect is to produce an ordering field at the ${\bf S}_3$ sites and order them in the direction 
opposite to the F-chain spins. The intra AF chain interaction orders the ${\bf S}_4$ spins  parallel to the F-chain spins, 
resulting in a 2D ferrimagnetic ground state. 
If $\eta_2 \ll \eta_3$ then the system will be more 2D, $m_1=m_2\cong 0.5$, and $m_3, m_4$ will be non-zero with the magnitude
of $m_3$ larger than $m_4$. On the other hand  if $\eta_2 \gg \eta_3$, then intra-chain AF bonds will dominate, 
making the AF chains nearly decoupled and the LRO in the AF chains will be small, $m_4 \approx -m_3 \ll 0.5$.  

\begin{figure}[httb]
\centering
\includegraphics[width=2.0in,clip]{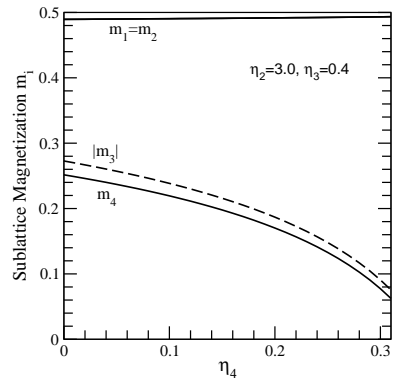}
\caption{Magnitude of sublattice magnetizations, $m_i$, for $\eta_2=3.0, \eta_3=0.4$ as function of $\eta_4$. 
Magnetizations for the two degenerate ($m_1=m_2$) ferro-modes (solid) corresponding to spins 1 and 2
slowly increase as $\eta_4$ is increased. Magnetizations $m_3, m_4$ for spins 3 (dashed) and 4 (solid)
decrease due to the increase in quantum fluctuations with increase in 
$\eta_4$. The ferrimagnetic ground state is stable in the parameter space $(\eta_2, \eta_3, \eta_4)$ 
as long as $\eta_3>\eta_4$ and $\eta_4\leq \eta_{4c}$, where $\eta_{4c}=0.316$ for $\eta_2=3.0, \eta_3=0.4$.}
\label{fig:magJ4}
\end{figure}

In Fig.~\ref{fig:magJ4}, we show 
how the ordered moments change 
with the increasing strength of the frustrated bond $\eta_4$ for specific values of $\eta_2=3.0$ and $\eta_3=0.4$. 
As $\eta_4$ approaches the critical value 0.316 the magnetization of the AF chain decreases but
remains finite ($|m_3| \sim 0.07, |m_4| \sim 0.06$) just before quantum phase transition to other ground state within LSWT. 
This is in contrast to what happens in the 
($J_1, J_2$) model where as $J_2$ approaches $J_c$ ($J_{c1}$ from the Ne\'{e}l state and $J_{c2}$ from the CAF state), the 
sublattice magnetization goes to zero.

\begin{figure}[httb]
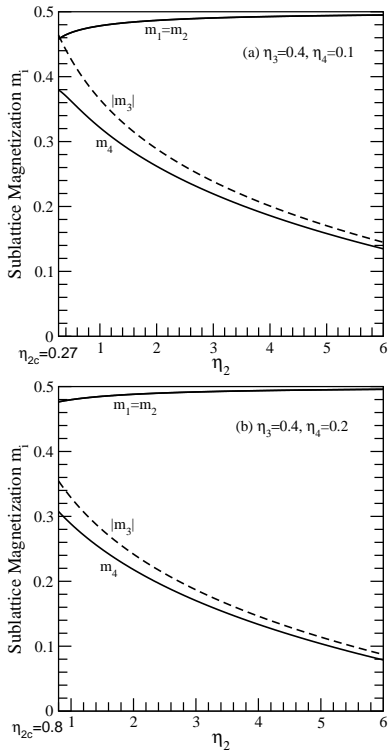

\centering
\includegraphics[width=2.0in,clip]{MagJ2-1.eps}
\qquad 
\includegraphics[width=2.0in,clip]{MagJ2-2.eps}
\caption{(a-b) Magnitude of sublattice magnetizations, $m_i$ for $\eta_3=0.4$, $\eta_4=0.1$ (Fig. a) and $0.2$ (Fig. b) 
as function of $\eta_2$. The ferrimagnetic ground state is stable for $\eta_2 \ge \eta_{2c}$ where $\eta_{2c}=0.27$
for $\eta_4=0.1$ and $\eta_2^c=0.80$ for $\eta_4=0.2$.
Ferro modes (solid) corresponding to spins 1 and 2 are degenerate ($m_1=m_2$). Magnetizations $m_3, m_4$ for spins 3 (dashed) 
and 4 (solid) decrease due to the increase in quantum fluctuations with increase in $\eta_2$.
}
\label{fig:magJ2}
\end{figure}

Finally, in Fig.~\ref{fig:magJ2}(a-b), we show the $\eta_2$ dependence of the magnitudes of the four order parameters $m_i$ ($i=1..4$) for  
$\eta_3=0.4$ for two fixed values of the frustrated interchain bond $\eta_4$. For our assumed collinear ferrimagnetic
ground state $\eta_3 > \eta_4$ and $\eta_2 > \eta_2^c$. For $\eta_3=0.4, \eta_4=0.1$, the critical value of $\eta_2^c$ is =0.27
and for $\eta_4=0.2$, $\eta_{2c}=0.80$.
For small $\eta_2$ i.e $\eta_2 \ll \eta_3$, $m_1=m_2=\vert m_3 \vert \sim 0.46$ and $m_4 \sim 0.38$, a reduction from 0.5 
by 8\% and 24\% respectively. The small antiferromagnetic coupling between spins of the AF-chain induces a relatively large value 
of the moment at the site 4. When $\eta_2$ increases the QSF in the AF-chain reduces the moments at sites 3 and 4. Notice that site 3 
still has a larger moment (in magnitude) than at site 4. For large $\eta_2$ values, say $\eta_2 \sim 6$, ferro chain spins 
have moments $\sim 0.495$,  whereas AF chain spins have moments of magnitude $\sim 0.14 (>0)$ due to small stabilizing 
interchain coupling $\eta_3=0.4$ [Fig.~\ref{fig:magJ2}(a)]. Increasing the strength of the frustrated bond $\eta_4$ essentially 
decouples the chains. For example with $\eta_4=0.2$, at $\eta_2=6.0$ ferro chains have moments close to 0.5 and AF-chains
have moments of magnitude $~0.08$ [Fig.~\ref{fig:magJ2}(b)]. 
For $\eta_2 < \eta_{2c}$, the system is most likely a spin liquid state without LRO.

\section{\label{sec:conclusions}Conclusions}
In summary, we have proposed a 2D frustrated Heisenberg model consisting of alternating 1D ferro ($J_1$)
and antiferro ($J_2$) chains which interact with alternating frustrated ($J_4$) and 
unfrustrated ($J_3$) bonds (strengths). The ground state is a long range ordered ferrimagnetic state in 
certain region of the parameter space. Analysis using linear spin wave theory suggests that the system undergoes 
a quantum phase transition to a quantum disordered phase with increasing strength of $\eta_4$, similar 
to the classic 2D ($J_1,J_2$) model. However in contrast to the ($J_1, J_2$) model, the sublattice magnetizations of the AF chains do not vanish at the critical 
value $\eta_{4c}$, similar to the 1D ${\rm S}_1=1, {\rm S}_2=1/2$ model of a quantum ferrimagnet. The exact nature of the 
phase transition, the nature of the GS above $\eta_{4c}$, and whether the order parameter vanishes at the transition should be explored by other theoretical 
and numerical techniques.

\section{Acknowledgment}
SDM would like to thank Dr. Xianglin Ke for stimulating discussions.
\appendix
\section{\label{EMM} Equation of Motion Method}
With inter-chain coupling (i.e. $\eta_3,\eta_4>0$), we are unable to find the the Bogoliubov transformations that diagonalizes 
the Hamiltonian in Eq.~\eqref{H0}. Thus we opt for another way - the canonical equation of motion method 
to obtain the magnon dispersion.\cite{diep} 
The various commutators that are needed for the canonical equation of motion method are:
\begin{widetext}
\begin{subequations}
\begin{eqnarray}
\big[a\k^{(1)}, {\cal H}_0/2J_1S\big] &=& A_1 a\k^{(1)}-B_1a\k^{(2)}+C_1b\k^{(3)\dag}+C_2^* a\k^{(4)},\label{comm1} \\
\big[ a\k^{(2)}, {\cal H}_0/2J_1S \big]&=& A_1 a\k^{(2)}-B_1a\k^{(1)}+C_1^*b\k^{(3)\dag}+C_2 a\k^{(4)}, \label{comm2}\\
 \big[a\k^{(4)}, {\cal H}_0/2J_1s\big] &=& A_2 a\k^{(4)}+B_2b\k^{(3)\dag}+C_2a\k^{(1)}+C_2^* a\k^{(2)},\label{comm3}\\
\big[b\k^{(3)\dag}, {\cal H}_0/2J_1s\big] &=& -A_3 b\k^{(3)\dag}-B_2a\k^{(4)}-C_1^*a\k^{(1)}-C_1 a\k^{(2)},\label{comm4}\\
\big[a\k^{(1)\dag}, {\cal H}_0/2J_1S\big] &=& -A_1 a\k^{(1)\dag}+B_1a\k^{(2)\dag}-C_1^*b\k^{(3)}-C_2 a\k^{(4)\dag}, \label{comm5}\\
\big[ a\k^{(2)\dag}, {\cal H}_0/2J_1S \big]&=& -A_1 a\k^{(2)\dag}+B_1a\k^{(1)\dag}-C_1b\k^{(3)}-C_2^* a\k^{(4)\dag}, \label{comm6}\\
 \big[a\k^{(4)\dag}, {\cal H}_0/2J_1s\big] &=& -A_2 a\k^{(4)\dag}-B_2b\k^{(3)}-C_2^*a\k^{(1)\dag}-C_2 a\k^{(2)\dag},\label{comm7}\\
\big[b\k^{(3)}, {\cal H}_0/2J_1s\big] &=& A_3 b\k^{(3)}+B_2a\k^{(4)\dag}+C_1a\k^{(1)\dag}+C_1^* a\k^{(2)\dag},\label{comm8}
\end{eqnarray}
\end{subequations}
\end{widetext}
where 
\begin{subequations}
\label{coeffs}
\bea
A_1 &=& (1+\eta_3-\eta_4),\; A_2=(\eta_2-2\eta_4),\\ 
A_3 &=& (\eta_2+2\eta_3),\;
B_1 = \gamma_y,\; B_2=\eta_2 \gamma_y,\\
C_1 &=& \eta_3 \gamma_xe^{ik_y b/4},\; C_2=\eta_4 \gamma_xe^{ik_y b/4}.
\eea
\end{subequations}
We notice
that the first four commutators [Eqs.~\eqref{comm1} -~\eqref{comm4}] are decoupled from the second four 
commutators [Eqs.~\eqref{comm5} -~\eqref{comm8}]. With the basis vectors $X\k=(a\k^{(1)}, a\k^{(2)}, a\k^{(4)}, b\k^{(3)\dag})$,
the canonical equation of motion can be deduced from the Hamiltonian in Eq.~\eqref{H0} in the following way:
\be
\Big[X\k^{(i)},\frac {{\cal H}_0}{2J_1S}\Big]=i\frac {dX\k^{(i)}}{dt}=g\omega\k^{(i)}X\k^{(i)}.
\label{EQC}
\ee
In Eq.~\eqref{EQC} $g$ is a $4 \times 4$ diagonal matrix with $g_{ii}=(1, 1, 1, -1)$ in the diagonal elements. 
The eigenvalues, $\omega\k^{(i)}$ are obtained by solving the determinant:
\be
\begin{vmatrix} (A_1 -\omega\k^{(i)}) & -B_1 & C_2 & C_1^* \\ 
-B_1 & (A_1 -\omega\k^{(i)}) & C_2^* & C_1 \\
C_2^* & C_2 & (A_2 -\omega\k^{(i)}) & B_2 \\
C_1 & C_1^* & B_2 & (A_3 +\omega\k^{(i)})
\end{vmatrix}=0.
\ee
The above determinant leads to a fourth-order polynomial:
\be
\omega\k^4+a\omega\k^3+b\omega\k^2+c\omega\k +d=0,
\label{poly}
\ee
where the coefficients are:
\begin{widetext}
\begin{subequations}
\bea
a &=& -2(1-2\eta_4), \label{coeffa}\\
b &=& (1+\eta_3-\eta_4)^2 -4(1+\eta_3-\eta_4)(\eta_3+\eta_4)-(\eta_2+2\eta_3)(\eta_2-2\eta_4)
-(1-\eta_2^2)\gamma_y^2+2(\eta_3^2-\eta_4^2)\gamma_x^2, \label{coeffb} \\
c&=& 2(1+\eta_3-\eta_4)^2(\eta_3+\eta_4)+2(1+\eta_3-\eta_4)(\eta_2-2\eta_4)(\eta_2+2\eta_3)
-2\eta_2^2(1+\eta_3-\eta_4)\gamma_y^2 -2(\eta_3+\eta_4)\gamma_y^2\non \\
&-&2(1+\eta_2+\eta_3-3\eta_4)\eta_3^2\gamma_x^2
+ 2(1-\eta_2-\eta_3-\eta_4)\eta_4^2 \gamma_x^2-2(\eta_3^2-\eta_4^2)\gamma_x^2\gamma_y^2+4\eta_2\eta_3\eta_4\gamma_x^2\gamma_y^2,
\label{coeffc} \\
d &=&-(1+\eta_3-\eta_4)\Big[(1+\eta_3-\eta_4)(\eta_2-2\eta_4)(\eta_2+2\eta_3)-(1+\eta_3-\eta_4)\eta_2^2\gamma_y^2 
- 2(\eta_2-2\eta_4)\eta_3^2\gamma_x^2 \non \\ 
&-&2(\eta_2+2\eta_3)\eta_4^2\gamma_x^2
+ 4\eta_2\eta_3\eta_4\gamma_x^2\gamma_y^2\Big]
+ (\eta_2-2\eta_4)(\eta_2+2\eta_3)\gamma_y^2+2(\eta_2-2\eta_4)\eta_3^2\gamma_x^2\gamma_y^2+2(\eta_2+2\eta_3)\eta_4^2\gamma_x^2\gamma_y^2\non \\
&-&\eta_2^2\gamma_y^4-4\eta_3^2\eta_4^2\gamma_x^2-4\eta_3\eta_4\gamma_x^2\gamma_y^2(\eta_2-\eta_3\eta_4).\label{coeffd}
\eea
\end{subequations}
\end{widetext}

The other set of four boson operators ($a_k^{(1)\dag}, a_k^{(2)\dag},a_k^{(4)\dag},b_k^{(3)}$) lead to 
a similar fourth order polynomial equation, but the signs before the linear and cubic terms are negative. 
There is thus a $\omega\k \leftrightarrow -\omega\k$ symmetry between the two sets of solutions.
This fourth order polynomial [Eq.~\eqref{poly}] has to be 
solved numerically. The four real eigen-values can be positive or negative.
If we solve the fourth order polynomial associated with the other four boson 
operators we will get again four real solutions which are negative of the solutions of Eq.~\eqref{poly}. For the 
magnon frequencies we will consider only the four positive solutions. The diagonalized quadratic Hamiltonian in 
terms of new basis vectors $\alpha\k^{(i=1..4)}$ becomes:
\be
{\cal H}_0=E_0+\sum_{{\bf k}\in {\rm BZ}} \sum_{i=1}^4\epsilon\k^{(i)}\alpha\k^{(i)\dag}\alpha\k^{(i)}.
\label{EMMH0}
\ee
$E_0$ contributes to the LSWT correction to the classical ground state energy.

For the special case $k_x=\pi/a$ i.e. $\gamma_x=0$ the solutions of the polynomial [Eq.~\eqref{poly}] 
can be obtained analytically. They are (we only consider the positive solutions):
\begin{subequations}\label{dispgx0}
\bea
&&\omega\k^{(1,2)}=(1+\eta_3-\eta_4) \pm \gamma_y, \\
&&\omega\k^{(3,4)}= (\eta_3+\eta_4) \pm \sqrt{(\eta_3-\eta_4+\eta_2)^2-\eta_2^2\gamma_y^2},\non \\
\eea
\end{subequations}
and thus the energies of the magnon spectrum are:
\begin{subequations}
\bea
&&\epsilon\k^{(1,2)}=2J_1S\big[(1+\eta_3-\eta_4) \pm \gamma_y\big],\\
&&\epsilon\k^{(3,4)}=2J_1 S\big\vert(\eta_3+\eta_4) \pm \sqrt{(\eta_3-\eta_4+\eta_2)^2-\eta_2^2\gamma_y^2}\big \vert.
\non \\
\eea
\end{subequations}
\bibliography{FAFChain}

\begin{thebibliography}{36}%
\makeatletter
\providecommand \@ifxundefined [1]{%
 \@ifx{#1\undefined}
}%
\providecommand \@ifnum [1]{%
 \ifnum #1\expandafter \@firstoftwo
 \else \expandafter \@secondoftwo
 \fi
}%
\providecommand \@ifx [1]{%
 \ifx #1\expandafter \@firstoftwo
 \else \expandafter \@secondoftwo
 \fi
}%
\providecommand \natexlab [1]{#1}%
\providecommand \enquote  [1]{``#1''}%
\providecommand \bibnamefont  [1]{#1}%
\providecommand \bibfnamefont [1]{#1}%
\providecommand \citenamefont [1]{#1}%
\providecommand \href@noop [0]{\@secondoftwo}%
\providecommand \href [0]{\begingroup \@sanitize@url \@href}%
\providecommand \@href[1]{\@@startlink{#1}\@@href}%
\providecommand \@@href[1]{\endgroup#1\@@endlink}%
\providecommand \@sanitize@url [0]{\catcode `\\12\catcode `\$12\catcode
  `\&12\catcode `\#12\catcode `\^12\catcode `\_12\catcode `\%12\relax}%
\providecommand \@@startlink[1]{}%
\providecommand \@@endlink[0]{}%
\providecommand \url  [0]{\begingroup\@sanitize@url \@url }%
\providecommand \@url [1]{\endgroup\@href {#1}{\urlprefix }}%
\providecommand \urlprefix  [0]{URL }%
\providecommand \Eprint [0]{\href }%
\providecommand \doibase [0]{http://dx.doi.org/}%
\providecommand \selectlanguage [0]{\@gobble}%
\providecommand \bibinfo  [0]{\@secondoftwo}%
\providecommand \bibfield  [0]{\@secondoftwo}%
\providecommand \translation [1]{[#1]}%
\providecommand \BibitemOpen [0]{}%
\providecommand \bibitemStop [0]{}%
\providecommand \bibitemNoStop [0]{.\EOS\space}%
\providecommand \EOS [0]{\spacefactor3000\relax}%
\providecommand \BibitemShut  [1]{\csname bibitem#1\endcsname}%
\let\auto@bib@innerbib\@empty
\bibitem [{\citenamefont {Diep}(2004)}]{diep}%
  \BibitemOpen
  \bibfield  {author} {\bibinfo {author} {\bibfnamefont {H.~T.}\ \bibnamefont
  {Diep}},\ }\href@noop {} {\emph {\bibinfo {title} {Frustrated Spin
  Systems}}},\ \bibinfo {edition} {1st}\ ed.\ (\bibinfo  {publisher} {World
  Scientific},\ \bibinfo {address} {Singapore},\ \bibinfo {year}
  {2004})\BibitemShut {NoStop}%
\bibitem [{\citenamefont {Lacroix}\ \emph {et~al.}(2011)\citenamefont
  {Lacroix}, \citenamefont {Mendels},\ and\ \citenamefont {Mila}}]{mila}%
  \BibitemOpen
  \bibfield  {author} {\bibinfo {author} {\bibfnamefont {C.}~\bibnamefont
  {Lacroix}}, \bibinfo {author} {\bibfnamefont {P.}~\bibnamefont {Mendels}}, \
  and\ \bibinfo {author} {\bibfnamefont {F.}~\bibnamefont {Mila}},\ }\href@noop
  {} {\emph {\bibinfo {title} {Introduction to Frustrated Magnetism}}},\
  \bibinfo {edition} {1st}\ ed.,\ Vol.\ \bibinfo {volume} {164}\ (\bibinfo
  {publisher} {Springer-Verlag},\ \bibinfo {address} {Berlin},\ \bibinfo {year}
  {2011})\BibitemShut {NoStop}%
\bibitem [{\citenamefont {Sachdev}(2001)}]{subir1}%
  \BibitemOpen
  \bibfield  {author} {\bibinfo {author} {\bibfnamefont {S.}~\bibnamefont
  {Sachdev}},\ }\href@noop {} {\emph {\bibinfo {title} {Quantum Phase
  Transitions}}},\ \bibinfo {edition} {1st}\ ed.\ (\bibinfo  {publisher}
  {Cambridge University Press},\ \bibinfo {address} {Cambridge, UK},\ \bibinfo
  {year} {2001})\BibitemShut {NoStop}%
\bibitem [{\citenamefont {Anderson}(1952)}]{anderson}%
  \BibitemOpen
  \bibfield  {author} {\bibinfo {author} {\bibfnamefont {P.~W.}\ \bibnamefont
  {Anderson}},\ }\href@noop {} {\bibfield  {journal} {\bibinfo  {journal}
  {Phys.\ Rev.}\ }\textbf {\bibinfo {volume} {86}},\ \bibinfo {pages} {694}
  (\bibinfo {year} {1952})}\BibitemShut {NoStop}%
\bibitem [{\citenamefont {Harris}\ \emph {et~al.}(1971)\citenamefont {Harris},
  \citenamefont {Kumar}, \citenamefont {Halperin},\ and\ \citenamefont
  {Hohenberg}}]{harris71}%
  \BibitemOpen
  \bibfield  {author} {\bibinfo {author} {\bibfnamefont {A.~B.}\ \bibnamefont
  {Harris}}, \bibinfo {author} {\bibfnamefont {D.}~\bibnamefont {Kumar}},
  \bibinfo {author} {\bibfnamefont {B.~I.}\ \bibnamefont {Halperin}}, \ and\
  \bibinfo {author} {\bibfnamefont {P.~C.}\ \bibnamefont {Hohenberg}},\
  }\href@noop {} {\bibfield  {journal} {\bibinfo  {journal} {Phys.\ Rev.\ B}\
  }\textbf {\bibinfo {volume} {3}},\ \bibinfo {pages} {961} (\bibinfo {year}
  {1971})}\BibitemShut {NoStop}%
\bibitem [{\citenamefont {Igarashi}(1993)}]{igar93}%
  \BibitemOpen
  \bibfield  {author} {\bibinfo {author} {\bibfnamefont {J.~I.}\ \bibnamefont
  {Igarashi}},\ }\href@noop {} {\bibfield  {journal} {\bibinfo  {journal} {J.
  Phys. Soc. Jpn.}\ }\textbf {\bibinfo {volume} {62}},\ \bibinfo {pages} {4449}
  (\bibinfo {year} {1993})}\BibitemShut {NoStop}%
\bibitem [{\citenamefont {Majumdar}(2010)}]{majumdar10}%
  \BibitemOpen
  \bibfield  {author} {\bibinfo {author} {\bibfnamefont {K.}~\bibnamefont
  {Majumdar}},\ }\href@noop {} {\bibfield  {journal} {\bibinfo  {journal}
  {Phys.\ Rev. B}\ }\textbf {\bibinfo {volume} {82}},\ \bibinfo {pages}
  {144407} (\bibinfo {year} {2010})}\BibitemShut {NoStop}%
\bibitem [{\citenamefont {Richter}\ \emph {et~al.}(2015)\citenamefont
  {Richter}, \citenamefont {Zinke},\ and\ \citenamefont {Farnell}}]{richter}%
  \BibitemOpen
  \bibfield  {author} {\bibinfo {author} {\bibfnamefont {J.}~\bibnamefont
  {Richter}}, \bibinfo {author} {\bibfnamefont {R.}~\bibnamefont {Zinke}}, \
  and\ \bibinfo {author} {\bibfnamefont {D.~J.~J.}\ \bibnamefont {Farnell}},\
  }\href@noop {} {\bibfield  {journal} {\bibinfo  {journal} {Eur. Phys. J. B}\
  }\textbf {\bibinfo {volume} {88}},\ \bibinfo {pages} {2} (\bibinfo {year}
  {2015})}\BibitemShut {NoStop}%
\bibitem [{\citenamefont {Bishop}\ \emph {et~al.}(2008)\citenamefont {Bishop},
  \citenamefont {Li}, \citenamefont {Darradi},\ and\ \citenamefont
  {Richter}}]{bishop}%
  \BibitemOpen
  \bibfield  {author} {\bibinfo {author} {\bibfnamefont {R.~F.}\ \bibnamefont
  {Bishop}}, \bibinfo {author} {\bibfnamefont {P.~H.~Y.}\ \bibnamefont {Li}},
  \bibinfo {author} {\bibfnamefont {R.}~\bibnamefont {Darradi}}, \ and\
  \bibinfo {author} {\bibfnamefont {J.}~\bibnamefont {Richter}},\ }\href@noop
  {} {\bibfield  {journal} {\bibinfo  {journal} {J.\ Phys.:\ Condens.\ Matter}\
  }\textbf {\bibinfo {volume} {20}},\ \bibinfo {pages} {255251} (\bibinfo
  {year} {2008})}\BibitemShut {NoStop}%
\bibitem [{\citenamefont {Sandvik}\ and\ \citenamefont
  {Singh}(2001)}]{sandvik01}%
  \BibitemOpen
  \bibfield  {author} {\bibinfo {author} {\bibfnamefont {A.~W.}\ \bibnamefont
  {Sandvik}}\ and\ \bibinfo {author} {\bibfnamefont {R.~R.~P.}\ \bibnamefont
  {Singh}},\ }\href@noop {} {\bibfield  {journal} {\bibinfo  {journal} {Phys.\
  Rev.\ Lett.}\ }\textbf {\bibinfo {volume} {86}},\ \bibinfo {pages} {528}
  (\bibinfo {year} {2001})}\BibitemShut {NoStop}%
\bibitem [{\citenamefont {Isaev}\ \emph {et~al.}(2009)\citenamefont {Isaev},
  \citenamefont {Ortiz},\ and\ \citenamefont {Dukelsky}}]{isaev}%
  \BibitemOpen
  \bibfield  {author} {\bibinfo {author} {\bibfnamefont {L.}~\bibnamefont
  {Isaev}}, \bibinfo {author} {\bibfnamefont {G.}~\bibnamefont {Ortiz}}, \ and\
  \bibinfo {author} {\bibfnamefont {J.}~\bibnamefont {Dukelsky}},\ }\href@noop
  {} {\bibfield  {journal} {\bibinfo  {journal} {Phys.\ Rev. B}\ }\textbf
  {\bibinfo {volume} {79}},\ \bibinfo {pages} {024409} (\bibinfo {year}
  {2009})}\BibitemShut {NoStop}%
\bibitem [{\citenamefont {Syromyatnikov}\ and\ \citenamefont
  {Aktersky}(2019)}]{syro}%
  \BibitemOpen
  \bibfield  {author} {\bibinfo {author} {\bibfnamefont {A.~V.}\ \bibnamefont
  {Syromyatnikov}}\ and\ \bibinfo {author} {\bibfnamefont {A.~Y.}\ \bibnamefont
  {Aktersky}},\ }\href@noop {} {\bibfield  {journal} {\bibinfo  {journal}
  {Phys.\ Rev. B}\ }\textbf {\bibinfo {volume} {99}},\ \bibinfo {pages}
  {224402} (\bibinfo {year} {2019})}\BibitemShut {NoStop}%
\bibitem [{\citenamefont {Yu}\ and\ \citenamefont {Kivelson}(2020)}]{kivelson}%
  \BibitemOpen
  \bibfield  {author} {\bibinfo {author} {\bibfnamefont {Y.}~\bibnamefont
  {Yu}}\ and\ \bibinfo {author} {\bibfnamefont {S.~A.}\ \bibnamefont
  {Kivelson}},\ }\href@noop {} {\bibfield  {journal} {\bibinfo  {journal}
  {Phys.\ Rev. B}\ }\textbf {\bibinfo {volume} {101}},\ \bibinfo {pages} {1580}
  (\bibinfo {year} {2020})}\BibitemShut {NoStop}%
\bibitem [{\citenamefont {Mikeska}\ and\ \citenamefont
  {Kolezhuk}(2008)}]{mikeska}%
  \BibitemOpen
  \bibfield  {author} {\bibinfo {author} {\bibfnamefont {H.-J.}\ \bibnamefont
  {Mikeska}}\ and\ \bibinfo {author} {\bibfnamefont {A.~K.}\ \bibnamefont
  {Kolezhuk}},\ }in\ \href@noop {} {\emph {\bibinfo {booktitle} {Quantum
  Magnetism, Lecture Notes in Physics}}},\ Vol.\ \bibinfo {volume} {645},\
  \bibinfo {editor} {edited by\ \bibinfo {editor} {\bibfnamefont
  {U.}~\bibnamefont {Schollw$\ddot{\rm o}$ck}}, \bibinfo {editor}
  {\bibfnamefont {J.}~\bibnamefont {Richter}}, \bibinfo {editor} {\bibfnamefont
  {D.~J.}\ \bibnamefont {Farnell}}, \ and\ \bibinfo {editor} {\bibfnamefont
  {R.~F.}\ \bibnamefont {Bishop}}}\ (\bibinfo  {publisher} {Springer},\
  \bibinfo {address} {Berlin, Heidelberg},\ \bibinfo {year} {2008})\ pp.\
  \bibinfo {pages} {1 -- 83}\BibitemShut {NoStop}%
\bibitem [{\citenamefont {Chubukov}\ \emph {et~al.}(1991)\citenamefont
  {Chubukov}, \citenamefont {Ivanova}, \citenamefont {Ivanov},\ and\
  \citenamefont {Korutcheva}}]{ivanova}%
  \BibitemOpen
  \bibfield  {author} {\bibinfo {author} {\bibfnamefont {A.~V.}\ \bibnamefont
  {Chubukov}}, \bibinfo {author} {\bibfnamefont {K.~I.}\ \bibnamefont
  {Ivanova}}, \bibinfo {author} {\bibfnamefont {P.~C.}\ \bibnamefont {Ivanov}},
  \ and\ \bibinfo {author} {\bibfnamefont {E.~R.}\ \bibnamefont {Korutcheva}},\
  }\href@noop {} {\bibfield  {journal} {\bibinfo  {journal} {J. Phys.: Condens.
  Matter}\ }\textbf {\bibinfo {volume} {3}},\ \bibinfo {pages} {2665} (\bibinfo
  {year} {1991})}\BibitemShut {NoStop}%
\bibitem [{\citenamefont {Kolezhuk}\ \emph {et~al.}(1997)\citenamefont
  {Kolezhuk}, \citenamefont {Mikeska},\ and\ \citenamefont
  {Yamamoto}}]{Kolezhuk1}%
  \BibitemOpen
  \bibfield  {author} {\bibinfo {author} {\bibfnamefont {A.~K.}\ \bibnamefont
  {Kolezhuk}}, \bibinfo {author} {\bibfnamefont {H.-J.}\ \bibnamefont
  {Mikeska}}, \ and\ \bibinfo {author} {\bibfnamefont {S.}~\bibnamefont
  {Yamamoto}},\ }\href@noop {} {\bibfield  {journal} {\bibinfo  {journal}
  {Phys.\ Rev. B}\ }\textbf {\bibinfo {volume} {55}},\ \bibinfo {pages} {R3336}
  (\bibinfo {year} {1997})}\BibitemShut {NoStop}%
\bibitem [{\citenamefont {Brehmer}\ \emph {et~al.}(1997)\citenamefont
  {Brehmer}, \citenamefont {Mikeska},\ and\ \citenamefont
  {Yamamoto}}]{brehmer}%
  \BibitemOpen
  \bibfield  {author} {\bibinfo {author} {\bibfnamefont {S.}~\bibnamefont
  {Brehmer}}, \bibinfo {author} {\bibfnamefont {H.-J.}\ \bibnamefont
  {Mikeska}}, \ and\ \bibinfo {author} {\bibfnamefont {S.}~\bibnamefont
  {Yamamoto}},\ }\href@noop {} {\bibfield  {journal} {\bibinfo  {journal} {J.
  Phys.: Condens. Matter}\ }\textbf {\bibinfo {volume} {9}},\ \bibinfo {pages}
  {3921} (\bibinfo {year} {1997})}\BibitemShut {NoStop}%
\bibitem [{\citenamefont {Pati}\ \emph
  {et~al.}(1997{\natexlab{a}})\citenamefont {Pati}, \citenamefont {Ramasesha},\
  and\ \citenamefont {Sen}}]{pati1}%
  \BibitemOpen
  \bibfield  {author} {\bibinfo {author} {\bibfnamefont {S.~K.}\ \bibnamefont
  {Pati}}, \bibinfo {author} {\bibfnamefont {S.}~\bibnamefont {Ramasesha}}, \
  and\ \bibinfo {author} {\bibfnamefont {D.}~\bibnamefont {Sen}},\ }\href@noop
  {} {\bibfield  {journal} {\bibinfo  {journal} {Phys. Rev. B}\ }\textbf
  {\bibinfo {volume} {55}},\ \bibinfo {pages} {8894} (\bibinfo {year}
  {1997}{\natexlab{a}})}\BibitemShut {NoStop}%
\bibitem [{\citenamefont {Pati}\ \emph
  {et~al.}(1997{\natexlab{b}})\citenamefont {Pati}, \citenamefont {Ramasesha},\
  and\ \citenamefont {Sen}}]{pati2}%
  \BibitemOpen
  \bibfield  {author} {\bibinfo {author} {\bibfnamefont {S.~K.}\ \bibnamefont
  {Pati}}, \bibinfo {author} {\bibfnamefont {S.}~\bibnamefont {Ramasesha}}, \
  and\ \bibinfo {author} {\bibfnamefont {D.}~\bibnamefont {Sen}},\ }\href@noop
  {} {\bibfield  {journal} {\bibinfo  {journal} {J. Phys.: Condens. Matter}\
  }\textbf {\bibinfo {volume} {9}},\ \bibinfo {pages} {8707} (\bibinfo {year}
  {1997}{\natexlab{b}})}\BibitemShut {NoStop}%
\bibitem [{\citenamefont {Ivanov}(1998)}]{NBIvanov0}%
  \BibitemOpen
  \bibfield  {author} {\bibinfo {author} {\bibfnamefont {N.~B.}\ \bibnamefont
  {Ivanov}},\ }\href@noop {} {\bibfield  {journal} {\bibinfo  {journal} {Phys.
  Rev. B}\ }\textbf {\bibinfo {volume} {57}},\ \bibinfo {pages} {R14 024}
  (\bibinfo {year} {1998})}\BibitemShut {NoStop}%
\bibitem [{\citenamefont {Kolezhuk}\ \emph {et~al.}(1999)\citenamefont
  {Kolezhuk}, \citenamefont {Mikeska}, \citenamefont {Maisinger},\ and\
  \citenamefont {Schollw$\ddot{\rm o}$ck}}]{Kolezhuk2}%
  \BibitemOpen
  \bibfield  {author} {\bibinfo {author} {\bibfnamefont {A.~K.}\ \bibnamefont
  {Kolezhuk}}, \bibinfo {author} {\bibfnamefont {H.-J.}\ \bibnamefont
  {Mikeska}}, \bibinfo {author} {\bibfnamefont {K.}~\bibnamefont {Maisinger}},
  \ and\ \bibinfo {author} {\bibfnamefont {U.}~\bibnamefont {Schollw$\ddot{\rm
  o}$ck}},\ }\href@noop {} {\bibfield  {journal} {\bibinfo  {journal} {Phys.\
  Rev. B}\ }\textbf {\bibinfo {volume} {59}},\ \bibinfo {pages} {13 565}
  (\bibinfo {year} {1999})}\BibitemShut {NoStop}%
\bibitem [{\citenamefont {Yamamoto}\ \emph {et~al.}(1998)\citenamefont
  {Yamamoto}, \citenamefont {Fukui}, \citenamefont {Maisinger},\ and\
  \citenamefont {Schollw$\ddot{\rm o}$ck}}]{shoji1}%
  \BibitemOpen
  \bibfield  {author} {\bibinfo {author} {\bibfnamefont {S.}~\bibnamefont
  {Yamamoto}}, \bibinfo {author} {\bibfnamefont {T.}~\bibnamefont {Fukui}},
  \bibinfo {author} {\bibfnamefont {K.}~\bibnamefont {Maisinger}}, \ and\
  \bibinfo {author} {\bibfnamefont {U.}~\bibnamefont {Schollw$\ddot{\rm
  o}$ck}},\ }\href@noop {} {\bibfield  {journal} {\bibinfo  {journal} {J.
  Phys.: Condens. Matter}\ }\textbf {\bibinfo {volume} {10}},\ \bibinfo {pages}
  {11 033} (\bibinfo {year} {1998})}\BibitemShut {NoStop}%
\bibitem [{\citenamefont {Maisinger}\ \emph {et~al.}(1998)\citenamefont
  {Maisinger}, \citenamefont {Schollw$\ddot{\rm o}$ck}, \citenamefont
  {Brehmer}, \citenamefont {Mikeska},\ and\ \citenamefont {Yamamoto}}]{shoji2}%
  \BibitemOpen
  \bibfield  {author} {\bibinfo {author} {\bibfnamefont {K.}~\bibnamefont
  {Maisinger}}, \bibinfo {author} {\bibfnamefont {U.}~\bibnamefont
  {Schollw$\ddot{\rm o}$ck}}, \bibinfo {author} {\bibfnamefont
  {S.}~\bibnamefont {Brehmer}}, \bibinfo {author} {\bibfnamefont {H.-J.}\
  \bibnamefont {Mikeska}}, \ and\ \bibinfo {author} {\bibfnamefont
  {S.}~\bibnamefont {Yamamoto}},\ }\href@noop {} {\bibfield  {journal}
  {\bibinfo  {journal} {Phys. Rev. B}\ }\textbf {\bibinfo {volume} {58}},\
  \bibinfo {pages} {R5908} (\bibinfo {year} {1998})}\BibitemShut {NoStop}%
\bibitem [{\citenamefont {Ivanov}(2000)}]{NBIvanov1}%
  \BibitemOpen
  \bibfield  {author} {\bibinfo {author} {\bibfnamefont {N.~B.}\ \bibnamefont
  {Ivanov}},\ }\href@noop {} {\bibfield  {journal} {\bibinfo  {journal} {Phys.
  Rev. B}\ }\textbf {\bibinfo {volume} {62}},\ \bibinfo {pages} {3271}
  (\bibinfo {year} {2000})}\BibitemShut {NoStop}%
\bibitem [{\citenamefont {Ivanov}(2009)}]{NBIvanov3}%
  \BibitemOpen
  \bibfield  {author} {\bibinfo {author} {\bibfnamefont {N.~B.}\ \bibnamefont
  {Ivanov}},\ }\href@noop {} {\bibfield  {journal} {\bibinfo  {journal}
  {Condensed Matter Physics}\ }\textbf {\bibinfo {volume} {12}},\ \bibinfo
  {pages} {435} (\bibinfo {year} {2009})}\BibitemShut {NoStop}%
\bibitem [{\citenamefont {Ivanov}\ and\ \citenamefont
  {Richter}(2001)}]{NBIvanov2}%
  \BibitemOpen
  \bibfield  {author} {\bibinfo {author} {\bibfnamefont {N.~B.}\ \bibnamefont
  {Ivanov}}\ and\ \bibinfo {author} {\bibfnamefont {J.}~\bibnamefont
  {Richter}},\ }\href@noop {} {\bibfield  {journal} {\bibinfo  {journal} {Phys.
  Rev. B}\ }\textbf {\bibinfo {volume} {63}},\ \bibinfo {pages} {144429}
  (\bibinfo {year} {2001})}\BibitemShut {NoStop}%
\bibitem [{\citenamefont {Lieb}\ and\ \citenamefont {Mattis}(1962)}]{Lieb}%
  \BibitemOpen
  \bibfield  {author} {\bibinfo {author} {\bibfnamefont {E.}~\bibnamefont
  {Lieb}}\ and\ \bibinfo {author} {\bibfnamefont {D.}~\bibnamefont {Mattis}},\
  }\href@noop {} {\bibfield  {journal} {\bibinfo  {journal} {J. Math. Phys.}\
  }\textbf {\bibinfo {volume} {3}},\ \bibinfo {pages} {749} (\bibinfo {year}
  {1962})}\BibitemShut {NoStop}%
\bibitem [{\citenamefont {Zhang}\ \emph {et~al.}(2020)\citenamefont {Zhang},
  \citenamefont {Zhao}, \citenamefont {Gautreau}, \citenamefont {Raczkowski},
  \citenamefont {Saha}, \citenamefont {Garlea}, \citenamefont {Cao},
  \citenamefont {Hong}, \citenamefont {Jeschke}, \citenamefont {Mahanti},
  \citenamefont {Birol}, \citenamefont {Assad},\ and\ \citenamefont {Ke}}]{Ke}%
  \BibitemOpen
  \bibfield  {author} {\bibinfo {author} {\bibfnamefont {H.}~\bibnamefont
  {Zhang}}, \bibinfo {author} {\bibfnamefont {Z.}~\bibnamefont {Zhao}},
  \bibinfo {author} {\bibfnamefont {D.}~\bibnamefont {Gautreau}}, \bibinfo
  {author} {\bibfnamefont {M.}~\bibnamefont {Raczkowski}}, \bibinfo {author}
  {\bibfnamefont {A.}~\bibnamefont {Saha}}, \bibinfo {author} {\bibfnamefont
  {V.~O.}\ \bibnamefont {Garlea}}, \bibinfo {author} {\bibfnamefont
  {H.}~\bibnamefont {Cao}}, \bibinfo {author} {\bibfnamefont {T.}~\bibnamefont
  {Hong}}, \bibinfo {author} {\bibfnamefont {H.~O.}\ \bibnamefont {Jeschke}},
  \bibinfo {author} {\bibfnamefont {S.~D.}\ \bibnamefont {Mahanti}}, \bibinfo
  {author} {\bibfnamefont {T.}~\bibnamefont {Birol}}, \bibinfo {author}
  {\bibfnamefont {F.~F.}\ \bibnamefont {Assad}}, \ and\ \bibinfo {author}
  {\bibfnamefont {X.}~\bibnamefont {Ke}},\ }\href@noop {} {\bibfield  {journal}
  {\bibinfo  {journal} {Phys.\ Rev. Lett.}\ }\textbf {\bibinfo {volume}
  {125}},\ \bibinfo {pages} {037204} (\bibinfo {year} {2020})}\BibitemShut
  {NoStop}%
\bibitem [{\citenamefont {Luttinger}\ and\ \citenamefont {Tisza}(1951)}]{LT}%
  \BibitemOpen
  \bibfield  {author} {\bibinfo {author} {\bibfnamefont {J.~M.}\ \bibnamefont
  {Luttinger}}\ and\ \bibinfo {author} {\bibfnamefont {L.}~\bibnamefont
  {Tisza}},\ }\href@noop {} {\bibfield  {journal} {\bibinfo  {journal} {Phys.\
  Rev.}\ }\textbf {\bibinfo {volume} {70}},\ \bibinfo {pages} {954} (\bibinfo
  {year} {1951})}\BibitemShut {NoStop}%
\bibitem [{\citenamefont {Lyons}\ and\ \citenamefont {Kaplan}(1960)}]{LK}%
  \BibitemOpen
  \bibfield  {author} {\bibinfo {author} {\bibfnamefont {D.~H.}\ \bibnamefont
  {Lyons}}\ and\ \bibinfo {author} {\bibfnamefont {T.~A.}\ \bibnamefont
  {Kaplan}},\ }\href@noop {} {\bibfield  {journal} {\bibinfo  {journal} {Phys.\
  Rev.}\ }\textbf {\bibinfo {volume} {120}},\ \bibinfo {pages} {1580} (\bibinfo
  {year} {1960})}\BibitemShut {NoStop}%
\bibitem [{\citenamefont {Igarashi}(1992)}]{igar92}%
  \BibitemOpen
  \bibfield  {author} {\bibinfo {author} {\bibfnamefont {J.~I.}\ \bibnamefont
  {Igarashi}},\ }\href@noop {} {\bibfield  {journal} {\bibinfo  {journal}
  {Phys.\ Rev. B}\ }\textbf {\bibinfo {volume} {46}},\ \bibinfo {pages} {10
  763} (\bibinfo {year} {1992})}\BibitemShut {NoStop}%
\bibitem [{\citenamefont {Igarashi}\ and\ \citenamefont
  {Nagao}(2005)}]{igar05}%
  \BibitemOpen
  \bibfield  {author} {\bibinfo {author} {\bibfnamefont {J.~I.}\ \bibnamefont
  {Igarashi}}\ and\ \bibinfo {author} {\bibfnamefont {T.}~\bibnamefont
  {Nagao}},\ }\href@noop {} {\bibfield  {journal} {\bibinfo  {journal} {Phys.\
  Rev. B}\ }\textbf {\bibinfo {volume} {72}},\ \bibinfo {pages} {014403}
  (\bibinfo {year} {2005})}\BibitemShut {NoStop}%
\bibitem [{\citenamefont {Holstein}\ and\ \citenamefont
  {Primakoff}(1940)}]{HP}%
  \BibitemOpen
  \bibfield  {author} {\bibinfo {author} {\bibfnamefont {T.}~\bibnamefont
  {Holstein}}\ and\ \bibinfo {author} {\bibfnamefont {H.}~\bibnamefont
  {Primakoff}},\ }\href@noop {} {\bibfield  {journal} {\bibinfo  {journal}
  {Phys.\ Rev.\ B}\ }\textbf {\bibinfo {volume} {58}},\ \bibinfo {pages} {1098}
  (\bibinfo {year} {1940})}\BibitemShut {NoStop}%
\bibitem [{\citenamefont {Bogoliubov}(1947)}]{Bogo}%
  \BibitemOpen
  \bibfield  {author} {\bibinfo {author} {\bibfnamefont {N.~N.}\ \bibnamefont
  {Bogoliubov}},\ }\href@noop {} {\bibfield  {journal} {\bibinfo  {journal} {J.
  Phys. (USSR)}\ }\textbf {\bibinfo {volume} {11}},\ \bibinfo {pages} {23}
  (\bibinfo {year} {1947})}\BibitemShut {NoStop}%
\bibitem [{\citenamefont {Colpa}(1978)}]{Colpa}%
  \BibitemOpen
  \bibfield  {author} {\bibinfo {author} {\bibfnamefont {J.}~\bibnamefont
  {Colpa}},\ }\href@noop {} {\bibfield  {journal} {\bibinfo  {journal}
  {Physica}\ }\textbf {\bibinfo {volume} {93A}},\ \bibinfo {pages} {327}
  (\bibinfo {year} {1978})}\BibitemShut {NoStop}%
\bibitem [{\citenamefont {Toth}\ and\ \citenamefont {Lake}(2015)}]{TL}%
  \BibitemOpen
  \bibfield  {author} {\bibinfo {author} {\bibfnamefont {S.}~\bibnamefont
  {Toth}}\ and\ \bibinfo {author} {\bibfnamefont {B.}~\bibnamefont {Lake}},\
  }\href@noop {} {\bibfield  {journal} {\bibinfo  {journal} {J. Phys.: Condens.
  Matter}\ }\textbf {\bibinfo {volume} {27}},\ \bibinfo {pages} {166002}
  (\bibinfo {year} {2015})}\BibitemShut {NoStop}%
\end{thebibliography}%

\end{document}